\documentclass[11pt]{article}
\usepackage{graphicx}
\usepackage{url}
\usepackage{makecell}
\usepackage{xcolor}
\usepackage{framed}
\usepackage{caption}
\usepackage{subcaption}
\usepackage{float}
\usepackage{textcomp}
\usepackage{hyperref}
\usepackage{amssymb}

\usepackage[format=plain,
            labelfont={bf,it},
            textfont={bf,it}]{caption}

\begin{document}

\thispagestyle{empty}

\begin{center}

Vrije Universiteit Amsterdam

\vspace{1mm}

\includegraphics[height=28mm]{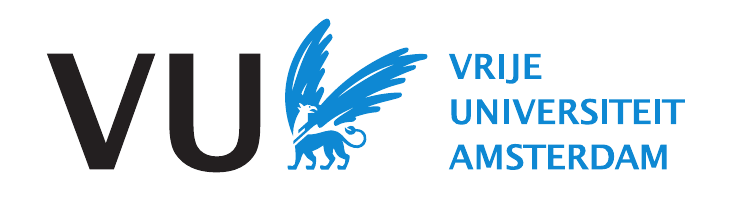}

\vspace{1.5cm}

{\Large Honours Program, Project Report}

\vspace*{1.5cm}

\rule{.9\linewidth}{.6pt}\\[0.4cm]
{\huge \bfseries Operational Characterization of a \\Public Scientific Datacenter During and \\Beyond the COVID-19 Period\par}
\rule{.9\linewidth}{.6pt}\\[1.5cm]

\vspace*{2mm}

{\Large
\begin{tabular}{l}
{\bf Author:} ~~Mehmet Berk Cetin~~~~ (2644886)
\end{tabular}
}

\vspace*{1.5cm}

\begin{tabular}{ll}
{\it 1st supervisor:}   & ~~prof. dr. ir. Alexandru Iosup \\
{\it daily supervisor:} & ~~ir. Laurens Versluis \\
{\it 2nd reader:}       & ~~1st supervisor \& daily supervisor
\end{tabular}

\vspace*{2cm}

\textit{A report submitted in fulfillment of the requirements for the Honours Program, \\ which is an excellence annotation to the VU Bachelor of Science degree in\\ Computer Science/Artificial Intelligence/Information Sciences\\
version 1.0}

\vspace*{1cm}

\today\\[4cm] 
\end{center}
\newpage


\newpage

\setcounter{tocdepth}{2}
\renewcommand*\contentsname{Table of Contents\newline}
\tableofcontents
\newpage

\section*{Abstract}
Datacenters are imperative for the digital society. They offer services such as computing, telecommunication, media, and entertainment. Datacenters, however, consume a lot of power. Improving datacenter operations is important and may result in better services, reduced energy consumption and reduced costs. To improve datacenters, we must understand what is going on inside them. Therefore, we use operational traces from a scientific cluster in the Netherlands to investigate and understand how that cluster operates.

Due to work-from-home circumstance, the covid period might have changed our daily usage of online applications, such as zoom and google meet. In this research, we focus on the operations of a scientific cluster~(LISA) inside the SURF datacenter. The global pandemic might have changed how the LISA cluster operates. To understand the change, we collect, combine, and analyze operational logs from the LISA cluster. The tool to collect the data that belongs to the non-covid period was accomplished in previous research. Nonetheless, both the tool and instrument to combine and analyze the traces are lacking. This research focuses on designing an instrument that can combine and analyze the traces during and before the coronavirus period. The instrument can also produce graphs for customarily selected rack, nodes and periods. Moreover, we characterize the traces that belong to the coronavirus period using the scientific instrument and additional tools. The outcome of this research helps us understand how the operations for a scientific cluster (LISA) in the Netherlands has changed after the global pandemic. \\ \\
\textbf{Keywords}~Datacenter; cluster; instrument; workload; operation; covid; non-covid;

\section{Introduction} \label{sec:introduction}
Datacenters are enormous systems that are designed to host thousands of computer servers, systems, and networking equipment that offer reliable and high speed services~\cite{journals/corr/IosupUVAEHTBT18}\cite{gens2014worldwide}. The digital society and economy is dependent on these systems to perform complex tasks and store data for future use. Datacenters, however, consume lots of energy. According to the research from Uptime Institute~\cite{bashroush2020beyond}, the annual power consumption of datacenters is between 400 terawatt-hours(TWh) and 500 TWh. Thus, for a better world, we must reduce the energy consumption and improve the services of datacenters. To achieve this, we must know how datacenters work. Unfortunately, knowing how datacenters operate isn't trivial, as they are like black boxes and we must obtain operational traces from the datacenter and investigate those traces in order to know how the datacenter operates. 

In this research, we design and develop a scientific instrument that can provide insights on the operations of a scientific cluster~(LISA) in the Netherlands. This instrument can also compare two different periods of the datacenter so we can understand how the datacenter operates in different time periods. For our research, the two different time periods is the covid and non-covid periods. From the LISA cluster, we have low-level operational traces that span approximately 7.5 months. The scientific instrument combines the low-level traces, performs basic statistical operations on them, and provides us visual information, mainly graphs, on how the cluster is operating. Moreover, this instrument is able to produce graphs for two different scenarios. First is the covid vs non-covid periods, as the instrument gives us two different graphs comparing the covid and non-covid periods. The second scenario is that the user can specify the analysis in a custom manner by passing arguments to the instrument. For instance, a system administrator can investigate two specific months for certain nodes and receive graphs and tables that help understand how those nodes operated, see Section~\ref{subsec:custom-analysis}.

In this research, we address the problem of not knowing how a scientific cluster in the Netherlands operates during the coronavirus period, and the lack of tools and instruments to find out the operations of the scientific cluster. Therefore, we design and develop the scientific instrument that can analyze and compare two different time periods~(mainly covid vs non-covid). In addition to that, to support data engineers and system administrators, the instrument can provide graphs and tables for node-level or rack-level operations, see Section~\ref{subsec:custom-analysis}.


The scientific instrument developed in this research is used to address the problem described in the previous paragraph. The main research question is how to design an instrument that can analyze the collected traces and statistically compare them during and before the coronavirus period. From the main research question, we derive three more research questions and describe the approach in section~\ref{sec:approach}.
This research has two main contributions. The first contribution is a scientific instrument that can give insight on how the LISA cluster operated per rack per node levels. The second contribution helps use understand how a scientific cluster in the Netherlands has operated during and before the coronavirus period, see Section~\ref{sec:approach}. Using different parameters, the instrument can produce graphs and tables for specific use cases. We believe the instrument can be useful to system administrators for performance evaluation in the past because the instrument works on data that has already been collected. 


\section{Background Information} \label{sec:background}

\textbf{SURF infrastructure} SURF is the biggest public scientific datacenter in the Netherlands. This datacenter is mostly used in scientific areas, such as physics, chemistry, machine learning, and computer systems. This datacenter has many clusters and one of them is the LISA cluster. Alex et al. has collected and open-sourced traces from the LISA cluster and they have the finest granularity~\cite{alex_uta_2020}. These traces span 3 months and they contain metrics that are gathered within a 15 second interval from over 300 computer nodes, where over 50 of the nodes are primarily used for running ML~(GPU nodes) workloads, which are called ML nodes and they mainly utilize their GPUs. The other 250+ nodes are used more to run Generic~(CPU nodes) workloads, which are called Generic nodes and mainly utilized their CPUs. In this research, we use these traces with an addition of 4.5 months. So, the date of the traces we use start from January 1 2020 and end in August 12 2020. 
\\ \\
\textbf{Terms \& definitions} A \textit{datacenter} is a giant building that hosts computer systems and its components. Inside datacenters, there may contain many \textit{clusters}. These clusters contain many computers that work together like a single computer system to accomplish various tasks. To understand how these clusters and datacenter work, we must collect \textit{operational traces} from these systems. The operational traces are high-level or low-level workload data collected either from the jobs and tasks run or from the utilized resources to run the jobs and tasks. \textit{High-level traces} contain job or task level data. In these high-level traces, there isn't much information on how the disk, memory, CPU, or GPU were utilized. We define \textit{low-level traces} as data that contains information on how the CPU, GPU, disk, and memory were utilized while running jobs and task. These traces also contain information on OS-level metrics such as network traffic and I/O operations.
\\ \\
\textbf{Updates to LISA} New ML nodes were added at the end of February and beginning March. We will consider this update while interpreting our graphs in Section~\ref{sec:results}.

\section{Methodology} \label{sec:methodology}
\subsection{Extracting, processing, and analyzing the data} \label{subsec:methodology:processing}
\begin{figure}[!t]
    \centering
    \includegraphics[width=\linewidth]{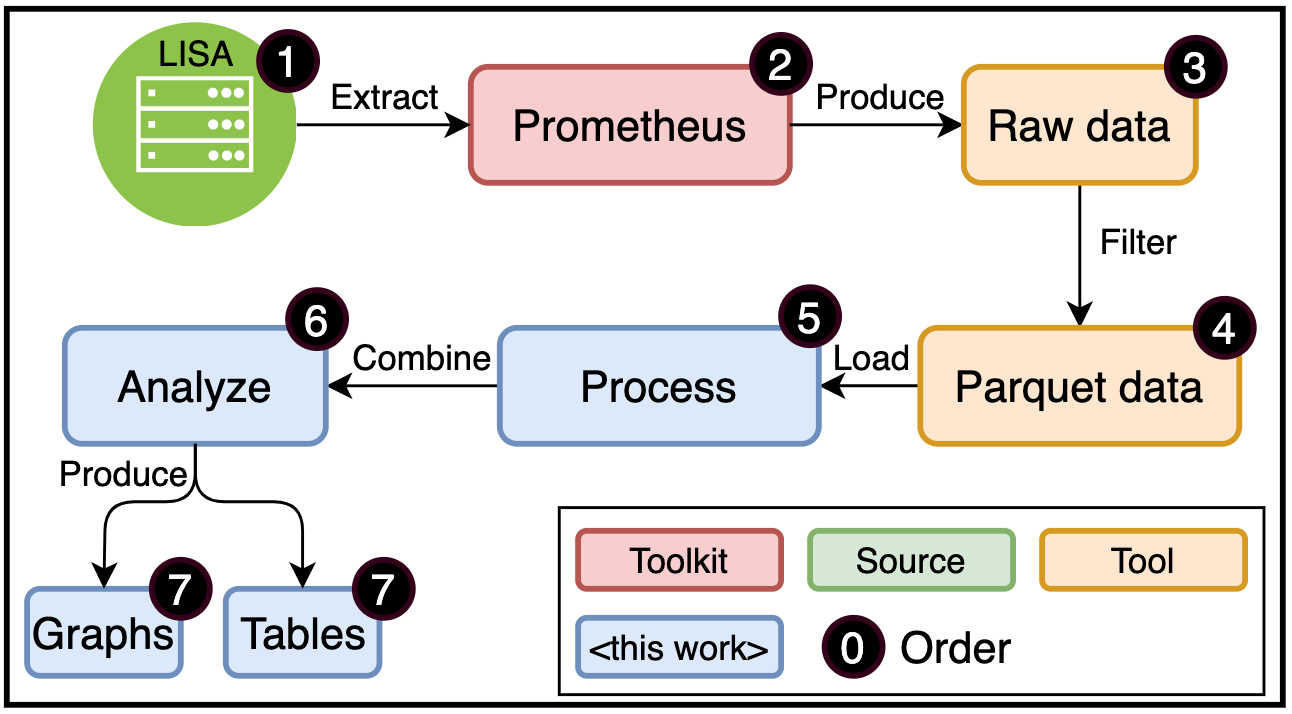}
    \caption{Data extraction, processing, and analysis.}
    \label{fig:extraction_model}
\end{figure}

The data used in this research was extracted from the LISA cluster using Prometheus as the toolkit. The whole workflow of extracting, processing, and analyzing the data is shown in Figure~\ref{fig:extraction_model}. Prometheus is an open-source toolkit that is used for systems monitoring and alerting toolkit originally built at SoundCloud~\cite{prometheus}. Moreover, Prometheus extracts relevant metrics from the system that it's monitoring and produces the data as raw data in JSON format. Afterwards, for faster processing, raw data is filtered and converted to parquet format. Next, the instrument processes the data according to user inputs (metric, period, rack, nodes). Last but not least, the instrument produces graphs and tables from the metrics that were processed earlier. Now, the user can interpret the results and see if there are any interesting insights on how the cluster performed.

The only part we did in Figure~\ref{fig:extraction_model} for this research is step 5, 6 and 7. The other steps were done by another research~\cite{alex_uta_2020}. To process and analyze the parquet data, we use Python 3.8.6, Pandas 1.1.0, Matplotlib 3.3.1, and Argparse 1.1. We first load the parquet data and process it based on user input. From the processed data, after the analysis, graphs and tables are produced. There are seven types of graphs, five of them are based on timestamps. Four of the graphs are daily and hourly graph. Two of daily and hourly graphs are daily seasonal and daily monthly graphs. The other two are hourly seasonal and hourly monthly graphs. The last graph is a curve showing the utilization of the metric across the whole season. Furthermore, the instrument produces cumulative distribution function graphs to show the distribution of the metric. For rack analysis, the instrument produces bar plots and violin plots to depict the distribution of the data. The tables show basic statistical values of the metric. 

\subsection{The Scientific Instrument} \label{subsec:methodology:instrument}
The instrument can produce graphs per node or per rack. If, however, the user doesn't pass a custom period and nodes or racks as arguments, all the nodes are processed by splitting the Generic and ML nodes and the timestamps are split into covid vs non-covid periods. Based on the user's desire, the instrument can produce various graphs for distinct metrics. For instance, the users of LISA are complaining to the system administrator and they claim that a rack or node isn't performing as expected. If system administrator wants to know the cause of this anomaly, the administrator, can use our instrument and pass arguments, one being the rack or node that isn't performing as expected and the other being the time period when the rack was down, to obtain graphs and tables on the performance of that rack or node.

\section{Approach \& Contributions} \label{sec:approach}





In this section, we describe the approach for each research question and mention the conceptual contributions. \\
To address {\bf RQ1} we design and develop an instrument that can statistically compare the operations of a cluster across different time periods. This instrument can split all the nodes by Generic and ML nodes. To address {\bf RQ2}, we use the method of using the scientific instrument and additional tools to interpret the covid period of the graphs and compare them to the non-covid period to understand the differences. We can understand how the covid period has effected the operations of the LISA cluster. To address {\bf RQ3}, we use the method of using the scientific instrument and additional tools to produce curves and violin plots, especially in rack level, for the covid and non-covid period. We interpret these curves and plots to characterize the operations for the covid period. The approaches should work because many other research papers~\cite{alex_uta_2020}\cite{amvrosiadis2018diversity}\cite{feitelson2014experience}\cite{iosup2008grid}\cite{shen2015statistical} use basic statistical methods together with curves, violin plots and bar plots. 

This research makes two main contributions and one potential contribution. The first contribution is a scientific instrument that can produce graphs for two different time periods for nodes that are selected based on the user's desire. The second contribution is the analysis for the covid and non-covid period. In terms of load1, power consumption, temperature, and RAM utilization, we saw that the covid period has effected how the LISA cluster operates. The third (potential) contribution is that this research encourages other scientists to collect, combine, and analyze traces from the LISA cluster for the post-covid period, which can be compared with before and during covid analysis.

\section{Results and Analysis} \label{sec:results}

The national Dutch Government declared a pandemic in 27 February 2020~\cite{corona}. From 1 January 2020 till the end of 26 February 2020, we refer the period as non-covid. From the beginning of 27 February 2020 to the end of the dataset, we refer the period as covid. 

In this section, we address all the research questions. Our main concern is to find out if the operations of the LISA cluster have changed during the covid period. Hence, we compare the general utilization levels, daily \& hourly utilization levels and rack utilization levels for the covid vs non-covid periods. Moreover, we showcase that any user of the instrument can select a custom node or rack and obtain graphs for any given time period.

\subsection{General Utilization Levels} \label{subsec:general-util}
\begin{figure}[H]
    \centering
    \includegraphics[width=\linewidth]{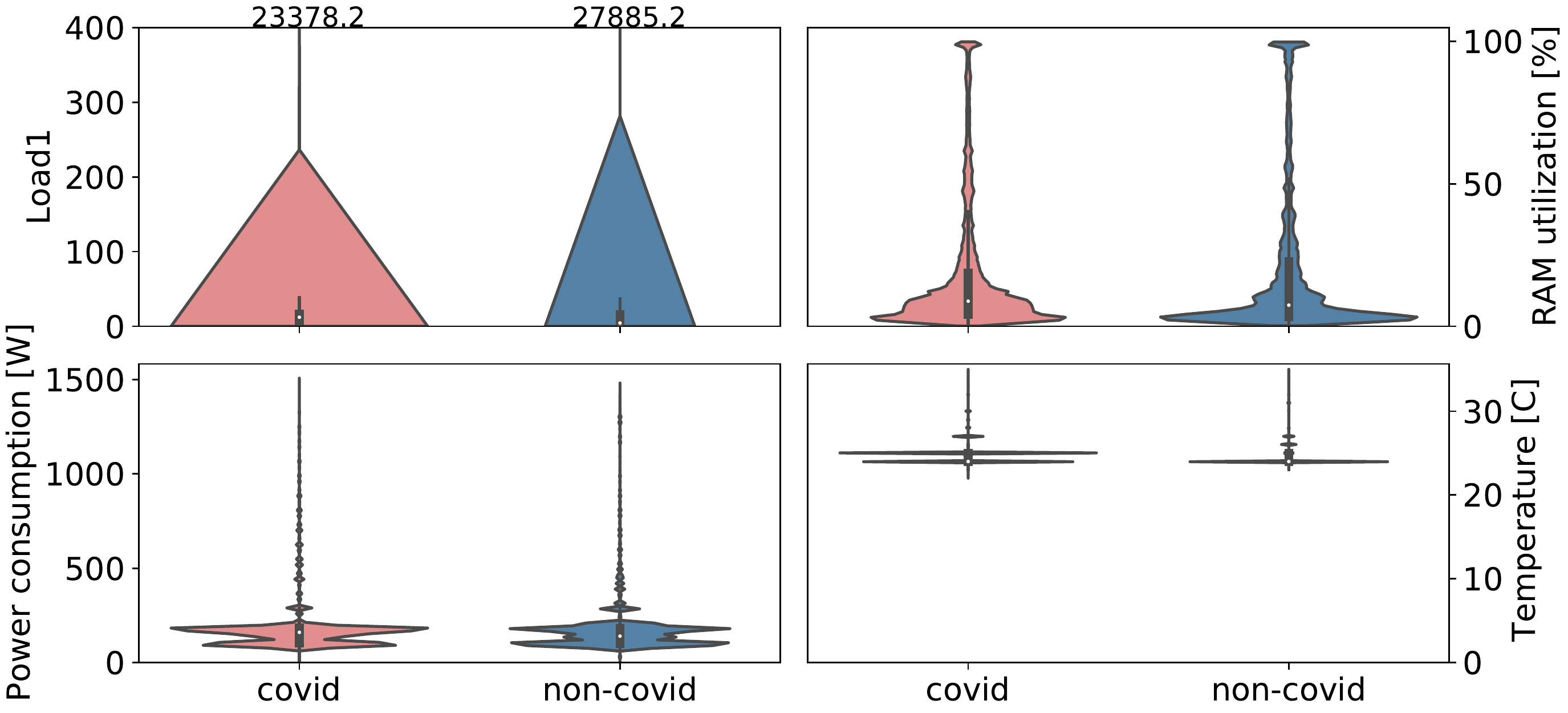}
    \caption{Load1, power consumption, RAM utilization and ambient temperature value distribution for the covid and non-covid periods across all the nodes.}
    \label{fig:covid-cluster-violinplots}
\end{figure}

\subsubsection{Power consumption}
In Figure~\ref{fig:covid-cluster-violinplots}, there seems to be no significant difference between the power consumption during and before the covid period. The IQR for both periods are almost same. Nonetheless, the upper outside points for the covid period are approximately 50W higher than the non-covid period. Furthermore, Figure~\ref{fig:covid-rack-barplots} indicates that all the ML node racks, except for Rack 33, consume 5W to 150W more average power during the covid period. Figure~\ref{fig:covid-rack-boxplots} depict the increase in power consumption during the covid period for the ML node racks. There is no significant difference in power consumption for the generic node racks. Nonetheless, Rack 23, consumes significantly more power, which peaks to approximately 1200W during the covid period. In Figure~\ref{fig:power-temp-r23}, the power consumption for Node 26 in Rack 23 increases from time to time, and peaks to approximately 1200W during the covid period.

\subsubsection{Load1}
In Figure~\ref{fig:covid-rack-boxplots}, almost all of the generic node racks have more instances where load1 values are higher during the covid period. Perhaps working from home has caused the user's of LISA (scientist \& engineers) to run intense jobs that hammer certain generic nodes or racks. These jobs might rapidly utilize the node's CPU and cause the load1 values to sky rocket for a certain amount of time.

In Figure~\ref{fig:covid-cluster-violinplots}, both periods contain nodes that have reached up to 23378 load. This indicates that there were times when the CPU's were highly utilized, some nodes were even hammered (see Section~\ref{subsec:rack-util}). The box plots for both periods have almost the same IQR. The upper whisker for both periods reach up to approximately 30 load. Nevertheless, the median (the white dot) for the covid period is higher than the non-covid period.

\subsubsection{Ambient Temperature}
\begin{figure}[H]
    \centering
    \includegraphics[width=\linewidth]{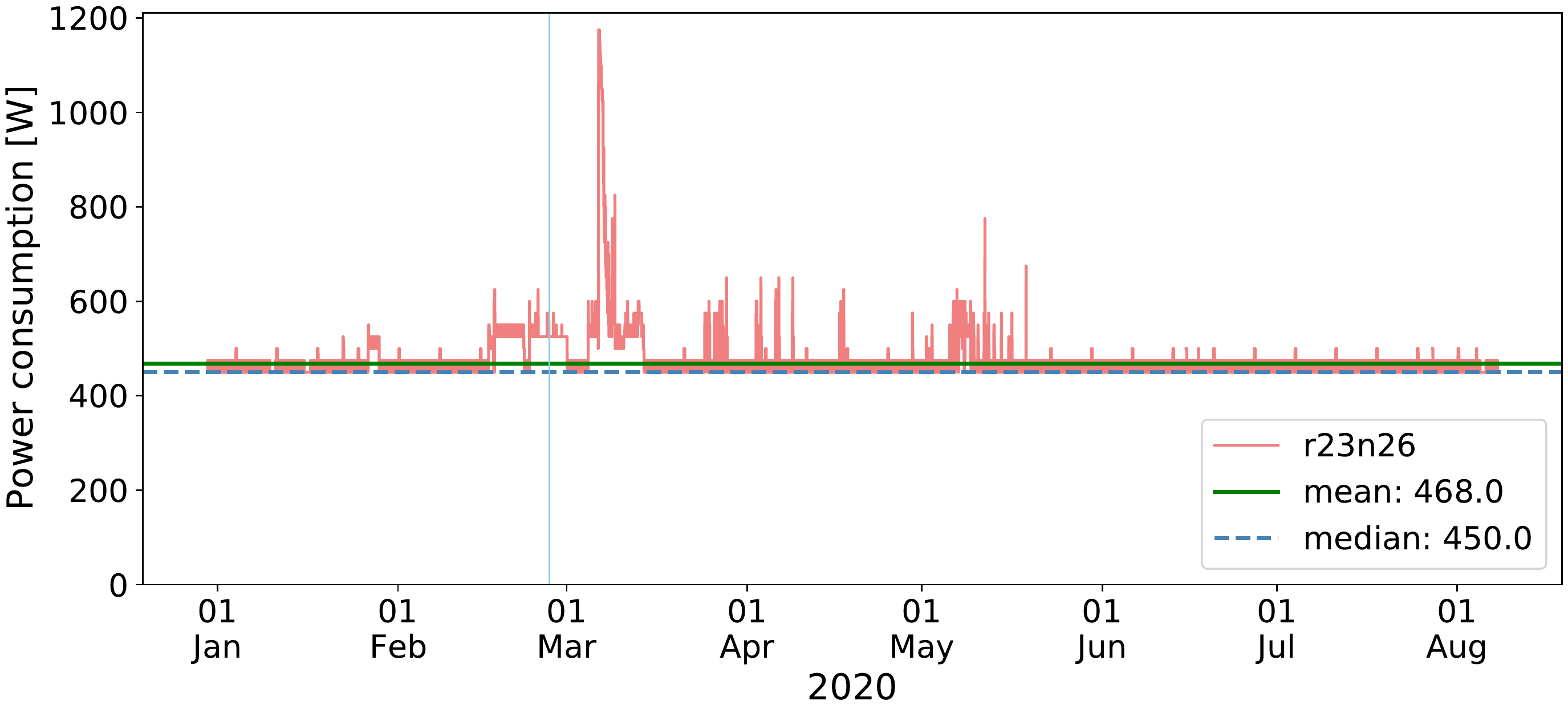}
    \includegraphics[width=\linewidth]{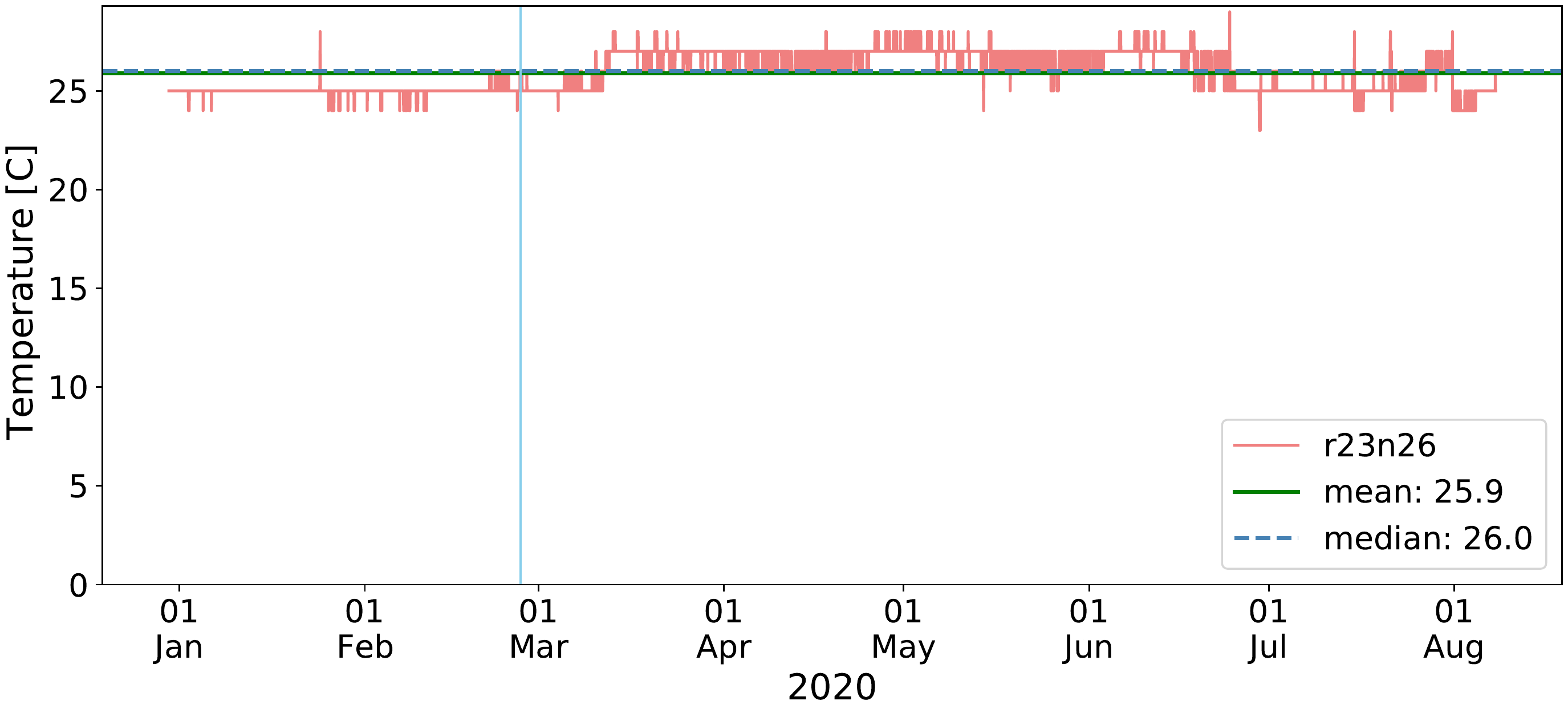}
    \caption{Temperature and power consumption for Rack 23 across the whole period. The vertical line separates the covid and non-covid periods.}
    \label{fig:power-temp-r23}
\end{figure}

The ambient temperature in Figure\ref{fig:covid-cluster-violinplots} depicts that the temperature for both periods are mostly 24\textdegree{}C. Nevertheless, the safe temperature for a server to work efficiently is 25\textdegree{}C. If servers work for a long time in an environment where the temperature is higher or lower than 25\textdegree{}C, the average failure probability for servers will be higher~\cite{pedram2012energy}. Moreover, in Figure~\ref{fig:covid-rack-barplots}, the average temperatures during the covid period for all ML node racks is 1\textdegree{}C to 3\textdegree{}C higher during the covid period. Out of all the generic node racks, only Rack 23 has, on average, higher temperature during the covid period. Figure~\ref{fig:power-temp-r23} depicts the 2\textdegree{}C average increase during the covid period for Node 26 in Rack 23.

\subsubsection{RAM Utilization}
According to the RAM utilization violin plot in Figure~\ref{fig:covid-cluster-violinplots}, the IQR is 10\% longer during the non-covid period, indicating a higher RAM utilization. Moreover, the upper outside points for the non-covid period is larger than the covid period, especially above 90\% RAM utilization. This means that there were more instances during the non-covid period where 90+\% RAM was utilized. Also, most of the generic node racks utilize more RAM during the non-covid period, see Section~\ref{subsec:rack-util}.

\subsection{Daily and Hourly Utilization Levels} \label{subsec:daily-hourly-util}
In this section, we investigate the power consumption and load1 values. Each graph gets the mean of the daily and hourly values, and aggregates them across all the Generic or ML nodes. 

\subsubsection{Power Consumption}
\begin{figure}[H]
    \centering
    \includegraphics[width=\linewidth]{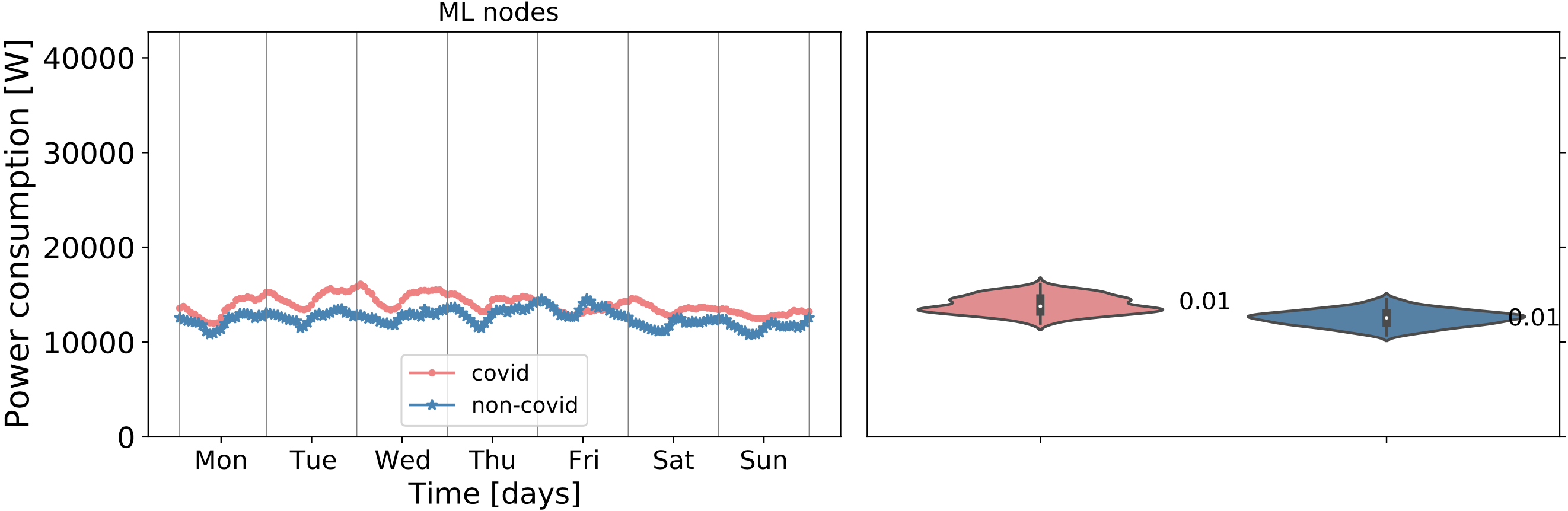}
    \caption{Daily power consumption for the covid and non-covid periods aggregated across all ML nodes. The maximum PDF value is annotated in the violin plot.}
    \label{fig:power-consumption-ml-daily}
\end{figure}
In Figure~\ref{fig:power-consumption-ml-daily}, the aggregated power consumption during the covid period for ML nodes is approximately 1000W higher than the non-covid period. There can be two reasons for this. The first one is that ML nodes were added to the LISA cluster at the end of February and beginning of March, see Section~\ref{sec:background}. The second reason can be that the cluster users have used LISA more during the covid period and that can lead to higher power consumption. 

\begin{figure}[H]
    \centering
    \includegraphics[width=\linewidth]{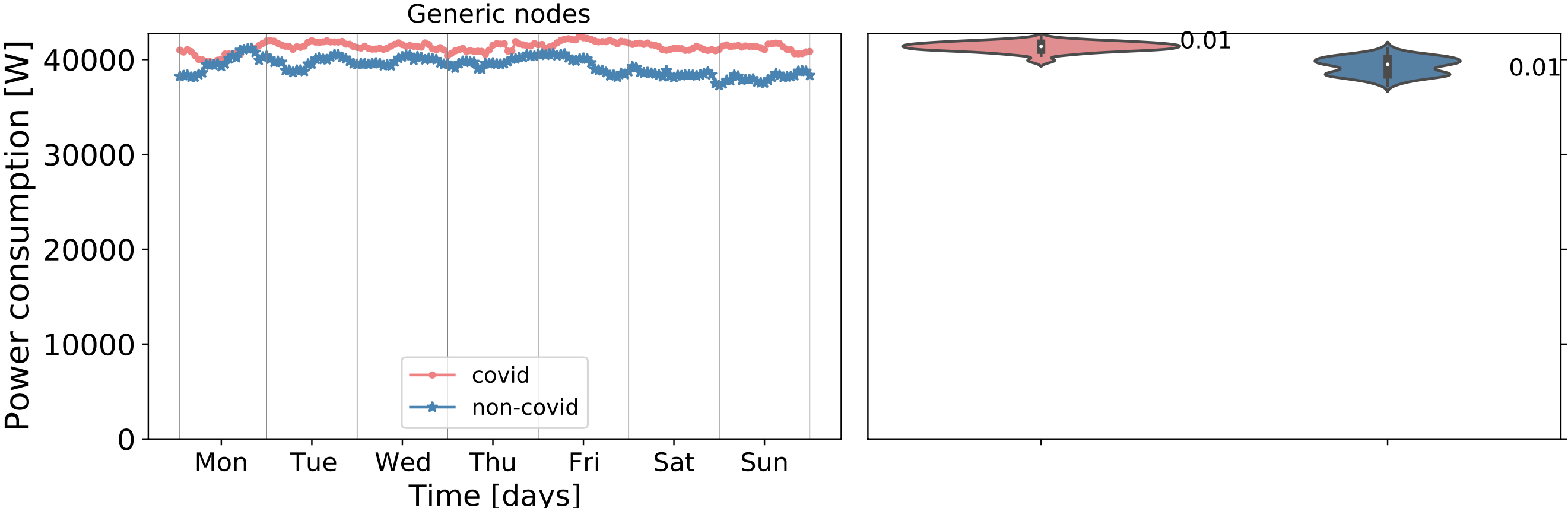}
    \caption{Daily power consumption for the covid and non-covid periods aggregated across all generic nodes. The maximum PDF value is annotated in the violin plot.}
    \label{fig:power-consumption-generic-daily}
\end{figure}
According to Figure~\ref{fig:power-consumption-generic-daily}, during the first 4 days of the week, the aggregated power consumption for the covid period is about 1500W higher than the non-covid period. After Friday noon, the gap between the two periods increase up to 2000W because the aggregated power consumption in the non-covid period decreases. The reason might be that cluster users no longer run jobs in the weekend because, during the non-covid period, people didn't work in the weekend. Perhaps, working from home in the covid period has caused people to work also during the weekend. 

\begin{figure}[H]
    \centering
    \includegraphics[width=\linewidth]{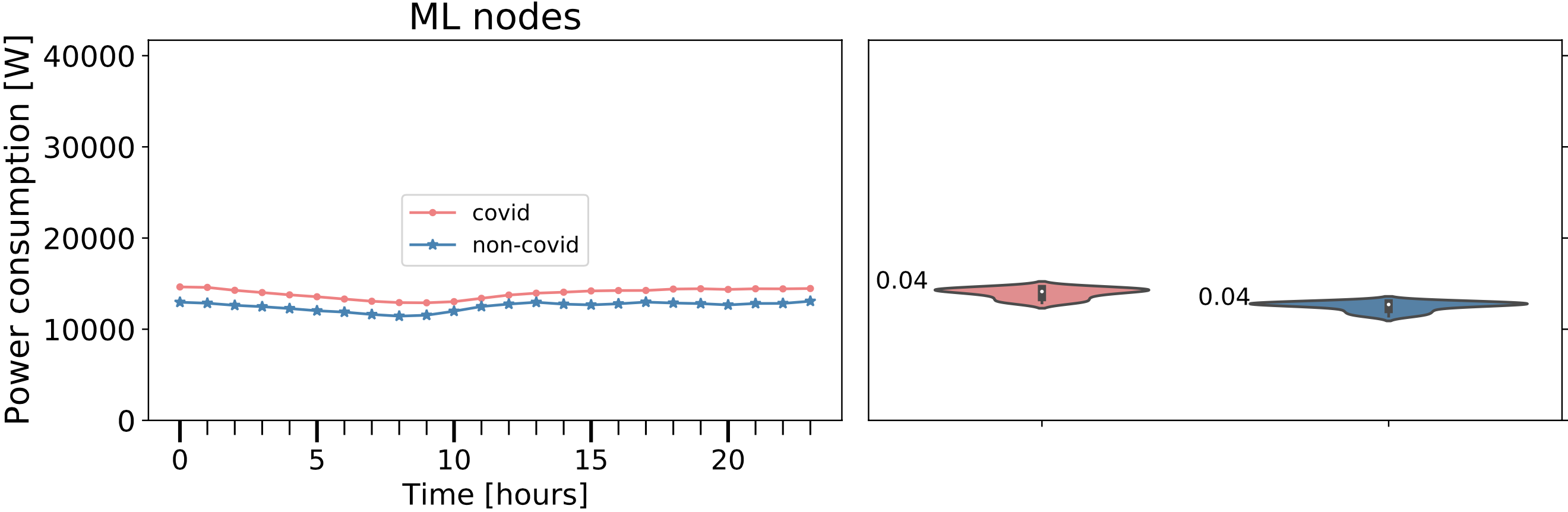}
    \caption{Hourly power consumption for the covid and non-covid periods aggregated across all ML nodes. The maximum PDF value is annotated in the violin plot.}
    \label{fig:power-consumption-ml-hourly}
\end{figure}
For the ML nodes, the curves for both periods are similar in Figure~\ref{fig:power-consumption-ml-hourly}. Both curves decrease during the night after 00:00 and increase after 9:00. Even if the curves for both periods are similar, throughout the day, the power consumption is approximately 1000 watts higher during the covid period. The cause of this can be the new added ML nodes (see Section~\ref{sec:background}), or the cluster users utilized the LISA more during the covid period. Both of the reasons can be the reason for the higher power consumption during the covid period.

\begin{figure}[H]
    \centering
    \includegraphics[width=\linewidth]{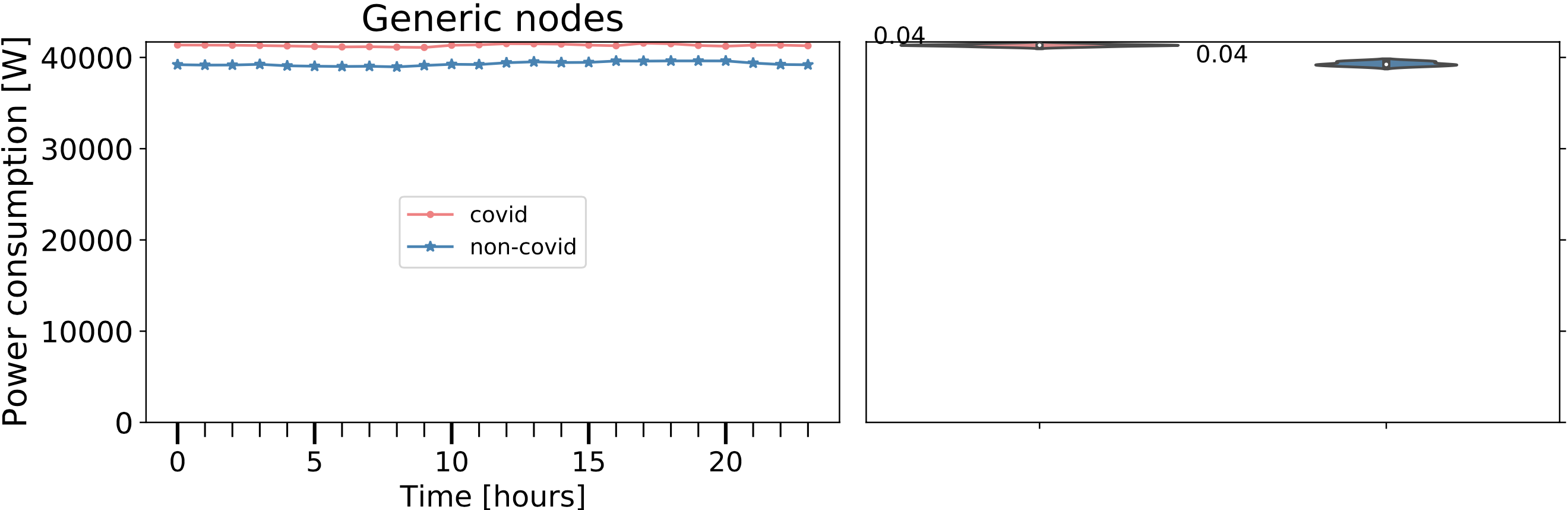}
    \caption{Hourly power consumption for the covid and non-covid periods aggregated across all generic nodes. The maximum PDF value is annotated in the violin plot.}
    \label{fig:power-consumption-generic-hourly}
\end{figure}
The power consumption during the covid period is around 1000W higher than the non-covid period in Figure~\ref{fig:power-consumption-generic-hourly}. Since there were no new generic nodes added to the cluster, the higher power consumption during the covid period can be caused by higher utilization levels. Moreover, for both periods, the power consumption for generic nodes don't vary throughout the day. Perhaps, the users of LISA run jobs also in the evening and night, or the users run long jobs that take days to finish. 

\subsubsection{Load}
\begin{figure}[H]
    \centering
    \includegraphics[width=\linewidth]{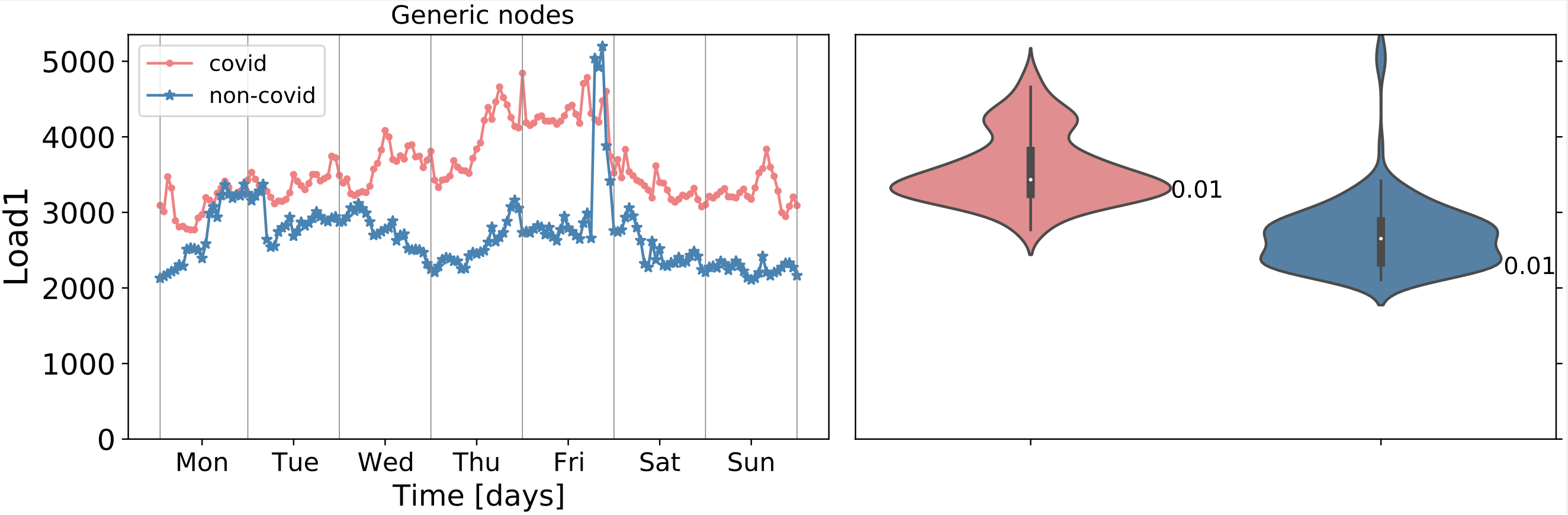}
    \caption{Daily load1 values for the covid and non-covid periods aggregated across all generic nodes. The maximum PDF value is annotated in the violin plot.}
    \label{fig:load1-generic-daily}
\end{figure}
In Figure~\ref{fig:load1-generic-daily}, during the covid period, aggregated load1 values are as high as 5000, especially during Thursday and Friday. There is only a moments where aggregated load1 is significantly higher during the non-covid period: Friday evening. Other than that, the power consumption is higher during the covid period. Furthermore, during the covid period, aggregated load1 values for Monday, Tuesday and Wednesday are almost the same as Saturday and Sunday. This indicates that cluster users run jobs also on weekends due to work from home policies. There is almost a 1000 load increase for both periods after Thursday morning and the values remain same throughout Friday. The reason can be that cluster users are trying to finish up their remaining tasks before the weekend.

\begin{figure}[H]
    \centering
    \includegraphics[width=\linewidth]{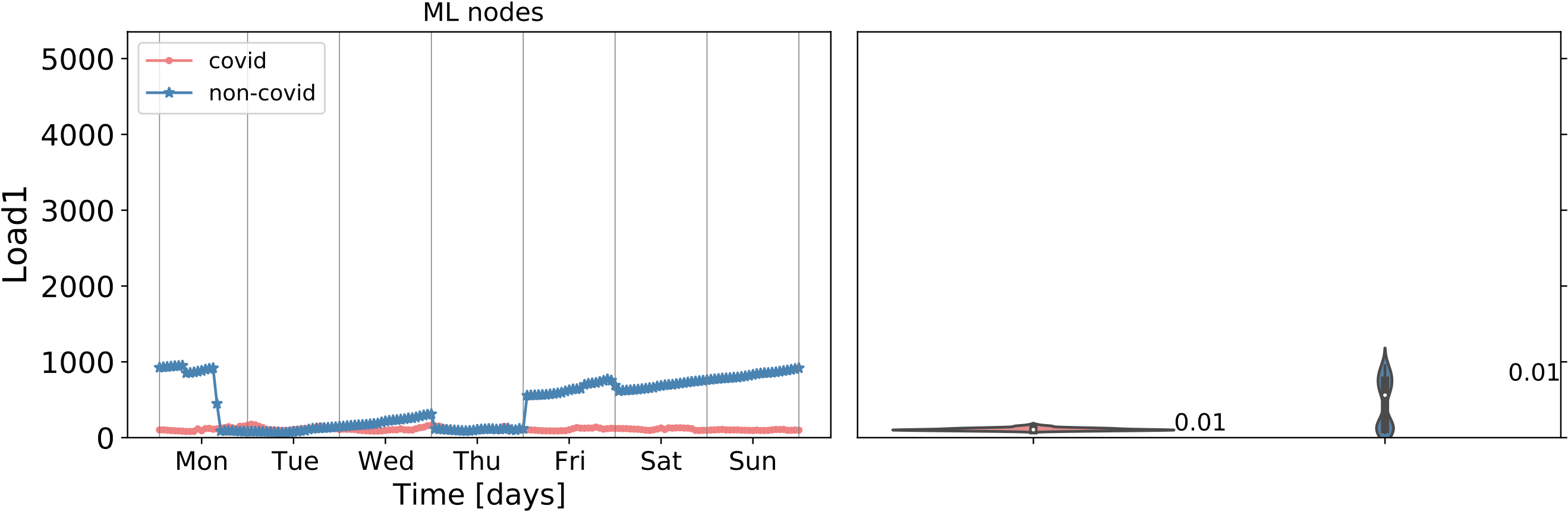}
    \caption{Daily load1 values for the covid and non-covid periods aggregated across all ML nodes. The maximum PDF value is annotated in the violin plot.}
    \label{fig:load1-ml-daily}
\end{figure}
As mentioned in Section~\ref{sec:background}, there were new ML nodes added at the end of February and the beginning of March. This didn't seem to have an effect to the covid period because load1 values don't change and they are close to zero, perhaps no more than 50. In the non-covid period, however, load1 values vary immensely.

\begin{figure}[H]
    \centering
    \includegraphics[width=\linewidth]{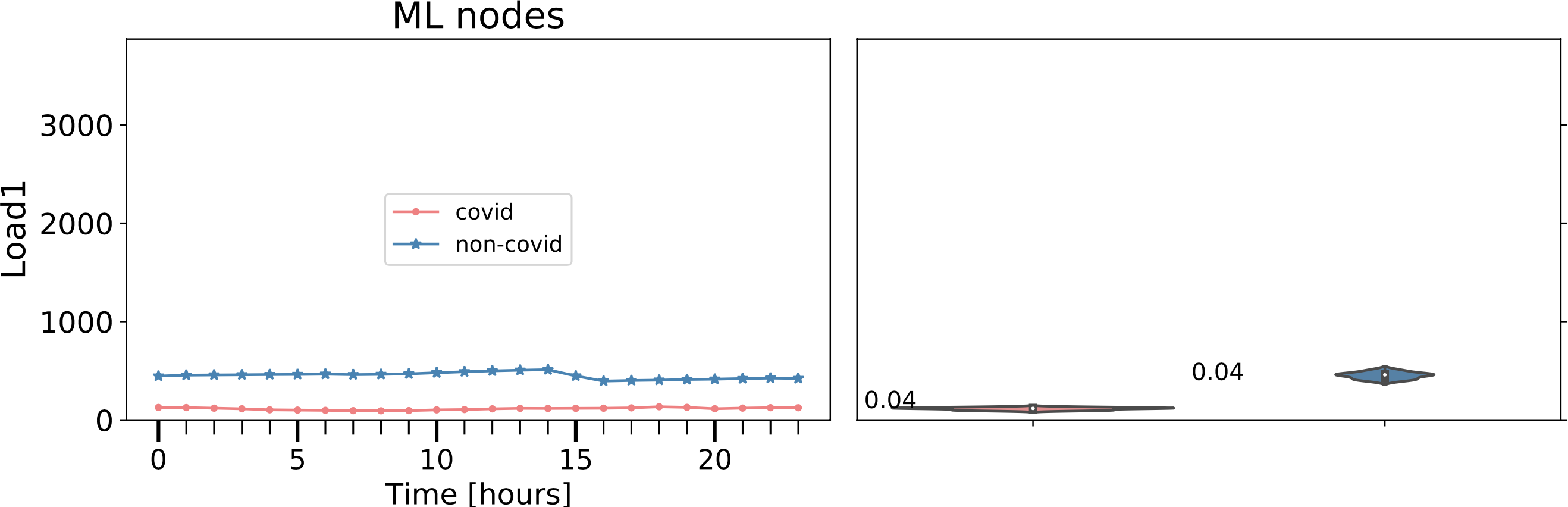}
    \caption{Hourly load1 values for the covid and non-covid periods aggregated across all ML nodes. The maximum PDF value is annotated in the violin plot.}
    \label{fig:load1-ml-hourly}
\end{figure}
In Figure~\ref{fig:load1-ml-hourly}, the load1 values for ML nodes during the covid period don't change throughout the day. The non-covid period is approximately 300 loads higher than the covid period. The reason might be that ML nodes don't utilize their CPU's much (because they run mainly ML workloads that utilize GPU) and that can cause the load1 values to remain low.

\begin{figure}[H]
    \centering
    \includegraphics[width=\linewidth]{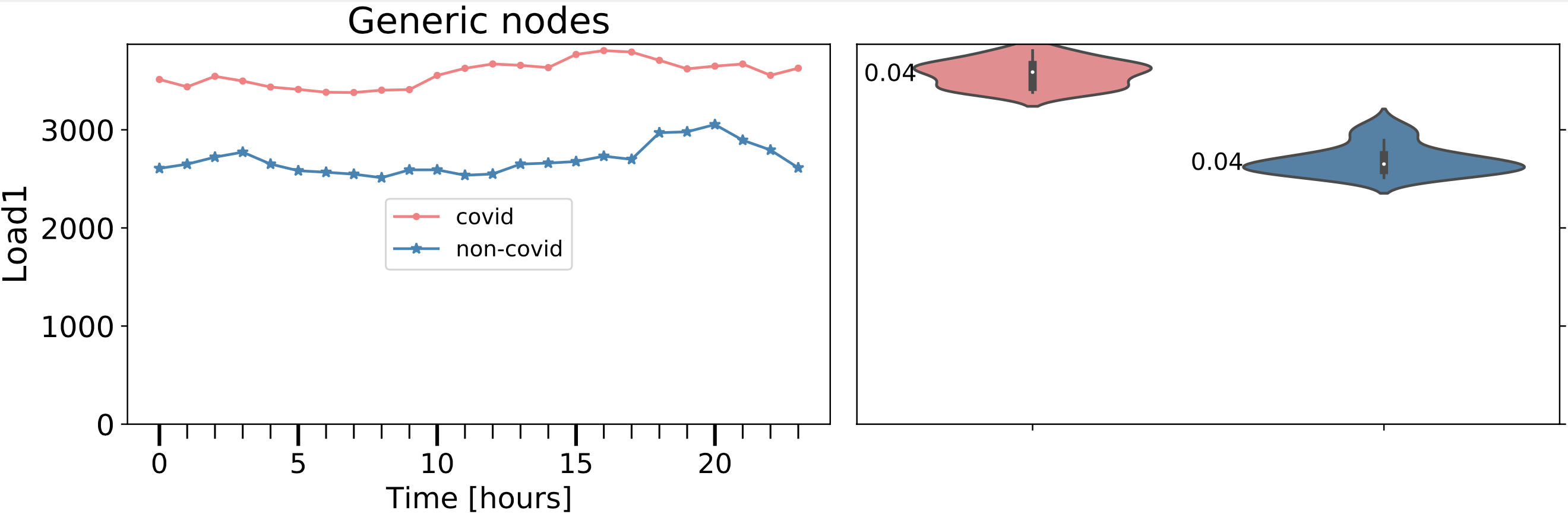}
    \caption{Hourly load1 values for the covid and non-covid periods aggregated across all generic nodes. The maximum PDF value is annotated in the violin plot.}
    \label{fig:load1-generic-hourly}
\end{figure}
generic nodes mainly run jobs that require CPU usage. Figure~\ref{fig:load1-generic-hourly} depicts that the generic nodes were utilized more during the covid period. The aggregated load1 values during the covid period are approximately 1000 loads higher than the non-covid period. This was almost the same in the daily load1 values for the generic nodes in Figure~\ref{fig:load1-generic-daily}. We can see that during the covid period, there is an increase in load1 values around 9:00, which is when people start working. After 17:00, the load1 values decrease for the covid period and increase for the non-covid period. The cause might be that the cluster users run their last jobs right before they leave the office, and those jobs took around 3 hours to finish (after 20:00 the load1 values for the non-covid period decrease).

\subsection{Rack Utilization levels} \label{subsec:rack-util}
In this section, we dive deeper into the cluster, as we investigate the periods in rack level. We compare all the racks between the covid and non-covid periods. 

\begin{figure}[H]
    \centering
    \includegraphics[width=\textwidth]{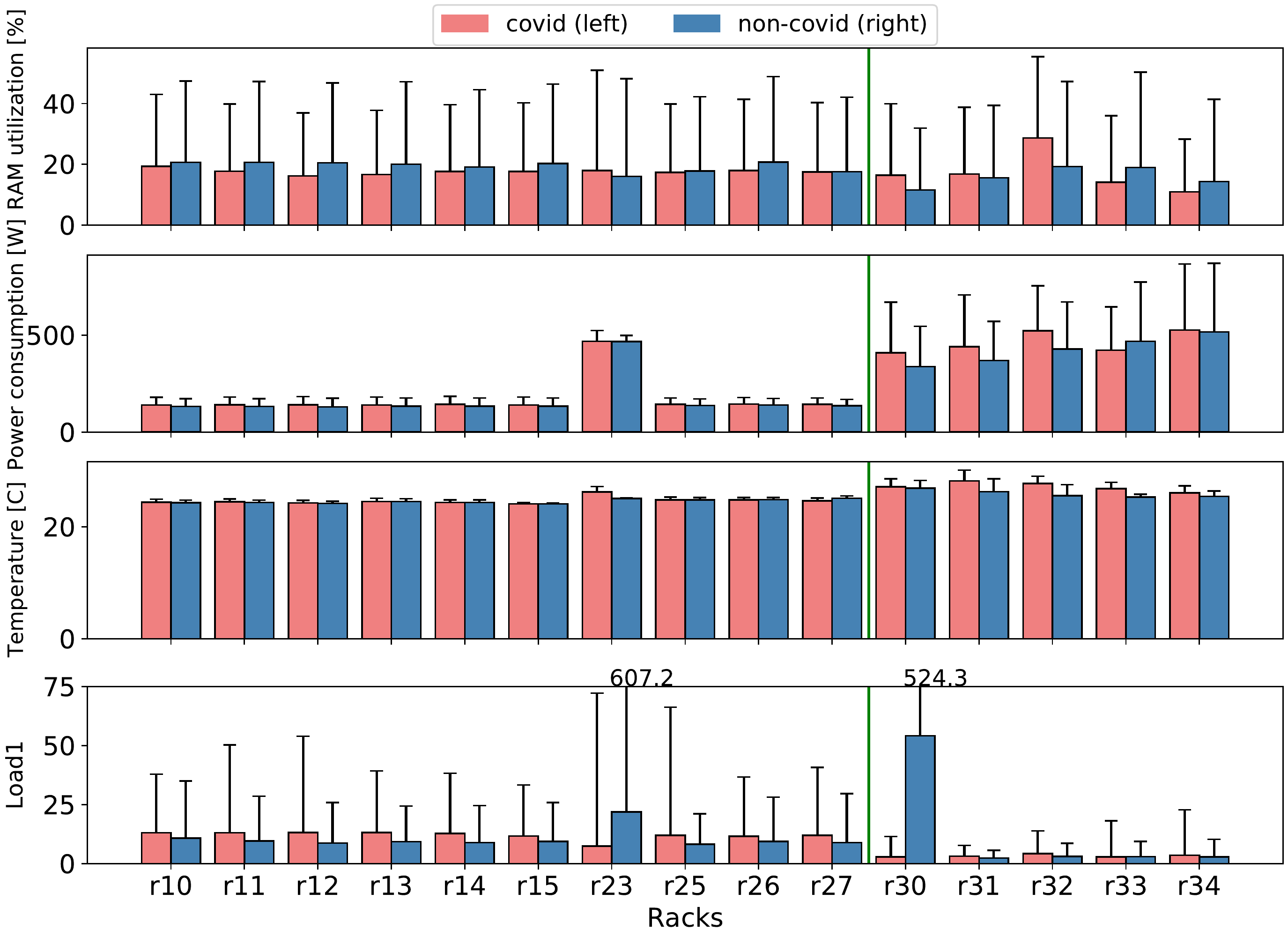}
    \caption{Average RAM utilization, load1, power consumption and ambient temperature values for each rack. The left side of the vertical line are the generic racks.}
    \label{fig:covid-rack-barplots}
\end{figure}
\begin{figure}[H]
    \centering
    \includegraphics[width=\textwidth]{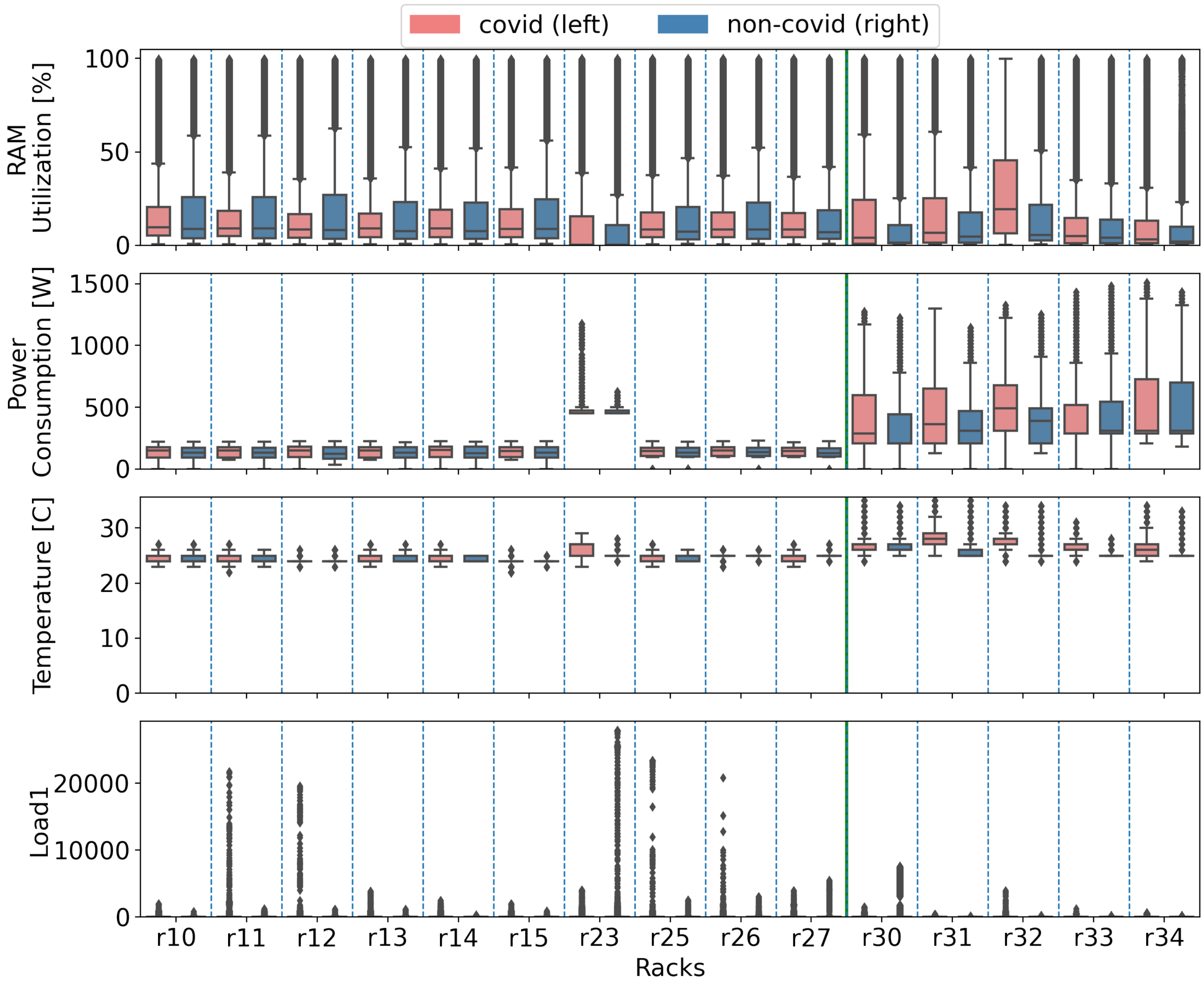}
    \caption{RAM utilization, load1, power consumption and ambient temperature value distribution for each rack. The left side of the green vertical line are the generic racks.}
    \label{fig:covid-rack-boxplots}
\end{figure}

\subsubsection{Power consumption}
In Figure~\ref{fig:covid-rack-barplots}, there isn't a significant difference in power consumption for the generic node racks. Nonetheless, according to Figure~\ref{fig:covid-rack-boxplots}, Rack 23 consume significantly more power during the covid period. The bar plots, however, don't depict Rack 23 much higher during the covid period because the average power consumption for Rack 23 remains the same for both periods. Nonetheless, during the covid period, the error bar for Rack 23 have a higher value, which means there are more outliers for that rack. This indicates that there were instances where up to 1200W power was consumed by Rack 23 during the covid period. The other generic node racks consume approximately 200W for both periods.

In Figure~\ref{fig:covid-rack-boxplots}, the ML node racks 30, 31, 32 and 34 consume more power during the covid period. Rack 33 consumes up to 1400W for both periods, which also indicates that ML node racks consume around 1000W-1200W more power than the generic node racks, where Rack 23 is an exception. In Figure~\ref{fig:covid-rack-barplots}, the ML node racks indicate that, on average, all the ML node racks, excluding r33, consume more power during the covid period. Moreover, these rack utilization levels reflect the results in Section~\ref{subsec:general-util} and~\ref{subsec:daily-hourly-util}, as the general, daily and hourly power consumption utilization levels were higher during the covid period, especially for ML node racks. Therefore, the power consumption during the covid period is higher than the non-covid period.

\subsubsection{Load1}
Figure~\ref{fig:covid-rack-barplots} depicts that during the covid period, the generic node racks have higher average load1 values. In Figure~\ref{fig:covid-rack-boxplots}, Rack 11, 12, 25 and 26 have moments where their load values were above 19500 during the covid period. This isn't same during the non-covid, as there is only Rack 23 that contains at least one load value over 19500. The bar plots depict that, during the covid period, all the generic  node racks, except for Rack 23, have higher load1 values. During the non-covid period, the violin plots depict that Rack 23 has the highest load1 value (27885). The bar plots show that, for the same rack and period, the average load1 value for Rack 23 is above 20 load and the standard deviation is 607.2.

The box plots for load1 depict that all the ML node racks, except for Rack 30 which has 7543 load as the highest load for the non-covid period, have higher load1 values during the covid period. Rack 30, 32 and 33 have moments where their load values were above 1000 during the covid period. For nodes that run primarily ML workloads, 1000 load is and extremely high load value because ML nodes mainly utilize their GPUs. Moreover, the violin plots depict that among all the racks, only the Racks 23, 27 and 30 have higher load1 values during the non-covid period. All the other racks have higher load1 values during the covid period. Perhaps, the reason for this is that users of the LISA cluster utilize the nodes more during the covid period.

\subsubsection{Ambient Temperature}
In Figure~\ref{fig:covid-rack-barplots}, all the generic node racks, except for Rack 23, have an average temperature close to 25°C for both periods. Rack 23 has 1\textdegree{}C to 2\textdegree{}C higher average temperature during the covid period. All the ML node racks have 1°C to 3°C higher average temperature during the covid period, especially Rack 31. In Figure~\ref{fig:covid-rack-boxplots}, all the ML node racks have significant outliers, which means high standard deviation, for both periods. Nonetheless, the generic node racks don't contain much outliers, indicating a low standard deviation. In general, the ML node racks have higher temperature from the generic node racks for both periods. The covid period for the ML nodes racks is hotter than their non-covid period.

\subsubsection{RAM Utilization}
In Figure~\ref{fig:covid-rack-barplots}, during the non-covid period, 8 Generic and 2 ML node racks approximately have 1\% to 5\% higher RAM utilization than the covid period. Rack 23 is the only Generic rack that has, on average, utilized (3\%) more RAM during the covid period. Across the whole cluster, the highest average RAM utilization occurs during the covid period by Rack 32. This rack has utilized its RAM on average 30\%, with a standard deviation exceeding 50\%. All the nodes, except for Rack 32, contain lots of outliers, see Figure~\ref{fig:covid-rack-boxplots}. This indicates a high standard deviation. The error bars in Figure~\ref{fig:covid-rack-barplots} depict that the RAM utilization in the LISA cluster has approximately 20\% standard deviation, meaning that the RAMs were utilized inconsistently high. 

\subsection{Overview of the COVID-19 period}
Overall, the covid period has caused some changes in the operations of the LISA cluster. The load1 values were significantly higher for certain racks during the covid period. Most of the racks consumed almost the same amount of power during the covid. Nonetheless, the power consumption during the covid period was slightly higher. Almost all of the generic node racks and some of the ML node racks have utilized their RAM more during the non-covid period. Average RAM utilization for all the racks was 10\% to 30\% and the standard deviation was about 20\%. The average ambient temperature for all the generic  node racks, except Rack 23, were the same for both periods. All the ML node racks have 1\textdegree{}C to 3\textdegree{}C higher temperature during the covid period. 

\subsection{Custom rack/node utilization levels} \label{subsec:custom-analysis}
\begin{figure}[H]
    \centering
    \includegraphics[width=\textwidth]{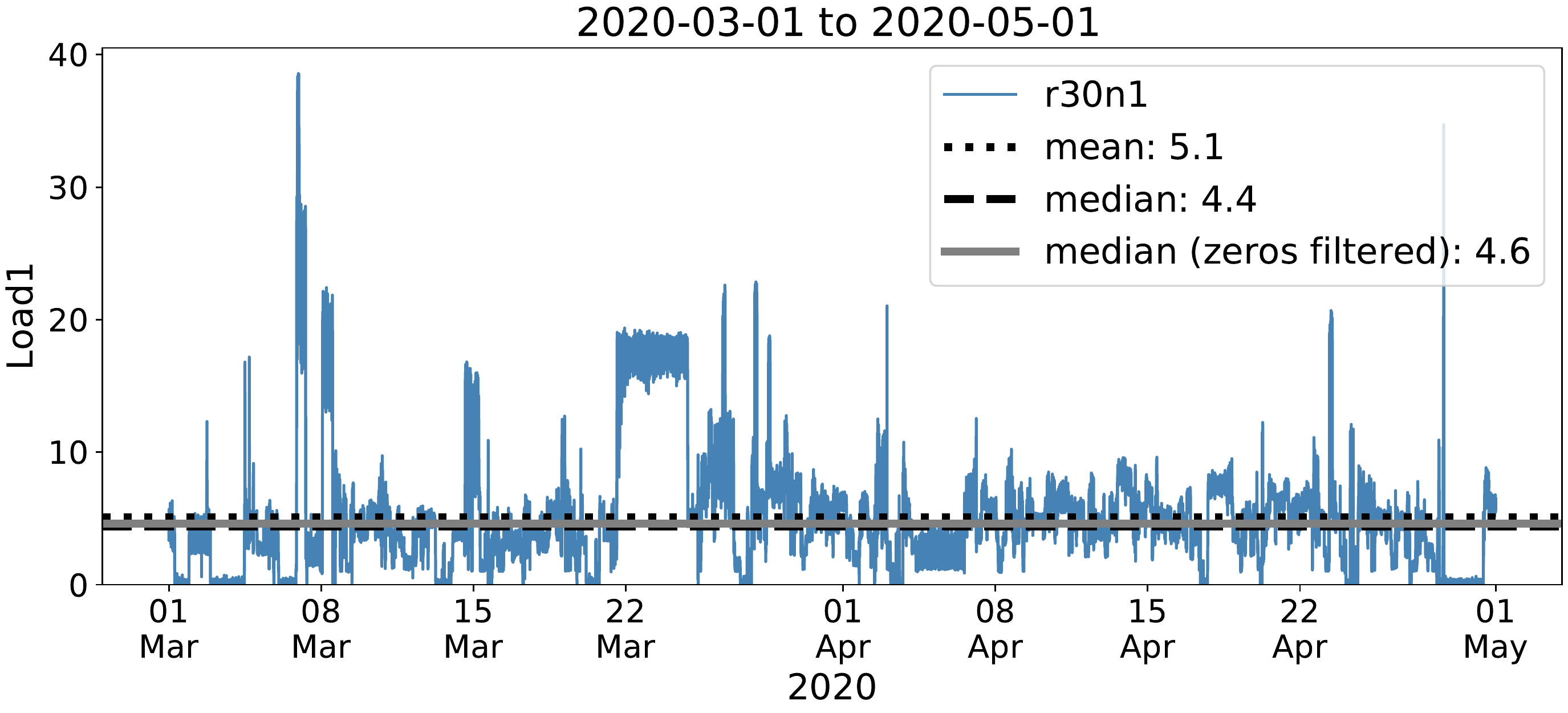}
    \includegraphics[width=\textwidth]{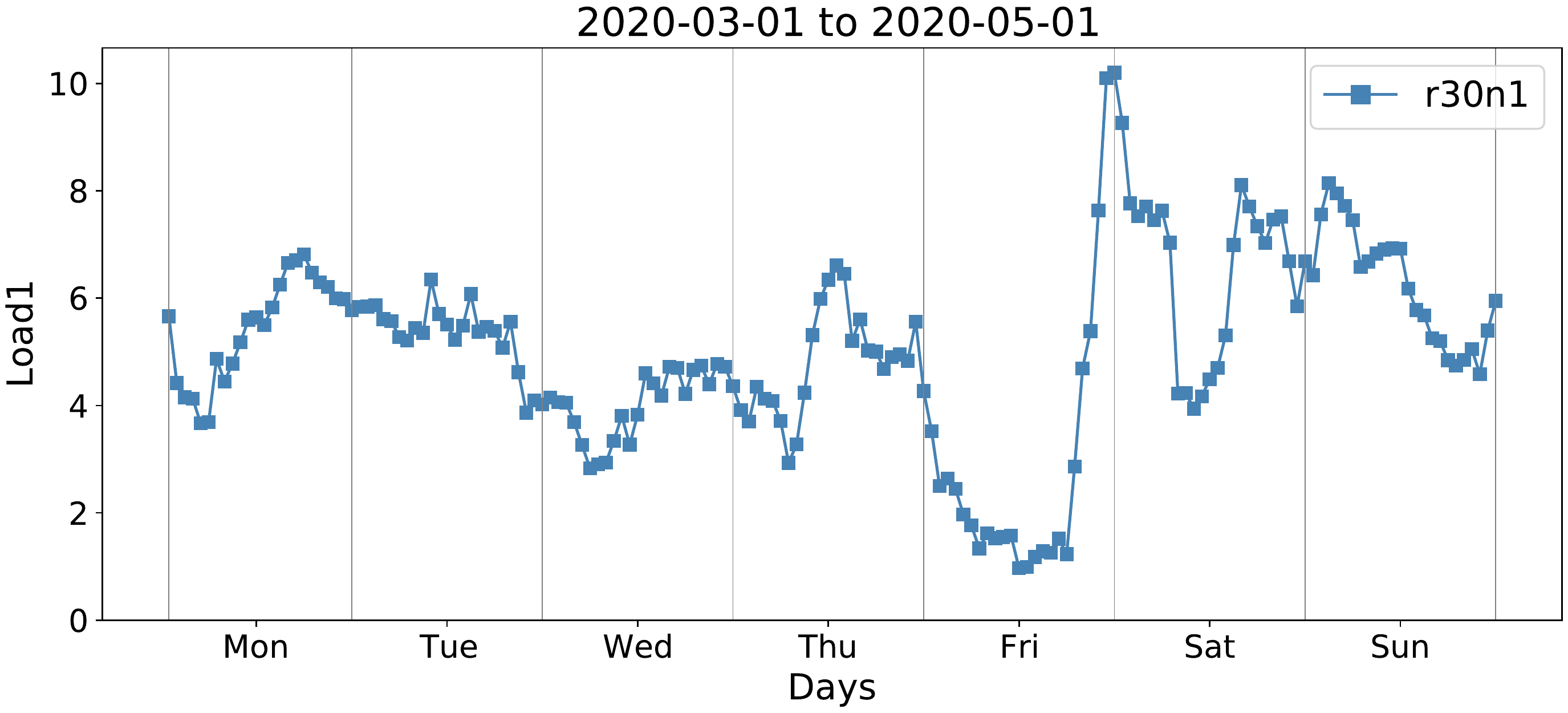}
    \includegraphics[width=\textwidth]{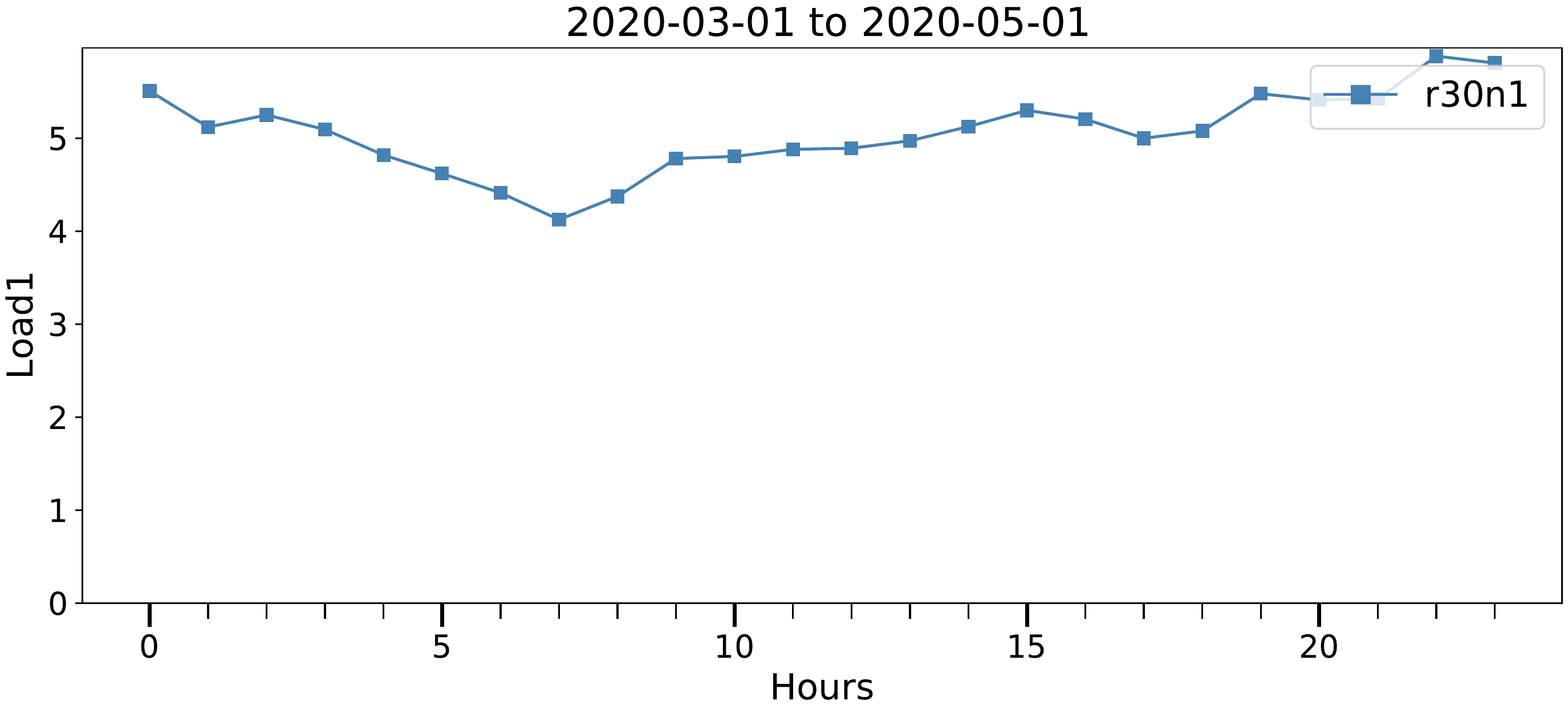}
    \caption{Daily, hourly, and seasonal load1 values for node r30n1.}
    \label{fig:node-load1}
\end{figure}

In this section, we explain the custom features of the scientific instrument. The instrument can produce graphs according to selected custom nodes, racks and periods, see Section~\ref{subsec:methodology:instrument}. The user of the instrument can specify a node or nodes or a rack, and a custom period. If a custom period isn't specified, the instrument will take the default periods (covid and non-covid). A data engineer or system administrator can use this instrument to delve deeper into the cluster to see how each rack or node is performing. For instance, if a system administrator wants to investigate node r30n1 from 1 March 2020 to 1 May 2020 for load1, the \nobreakspace administrator can input the parameters and the instrument produces graphs like Figure~\ref{fig:node-load1}. If that administrator also wants to compare load1 ƒor the same node with a different time period, the administrator can input the required parameters and the instrument produces graph similar graphs. The administrator can compare the performance of node r30n1 for two different time periods. The same comparison can be done for multiple nodes, one rack, or all nodes. 

\section{Threats to Validity} \label{sec:validity}

\textbf{The extracted dataset has some deficiencies, Prometheus might contain bugs}.
    Prometheus monitoring system is one of the most popular system-monitoring toolkit. This system is used by many scientists and engineers and we expect Prometheus to produce correct results. Prometheus, however, can contain bugs that affect the data collection mechanism. For instance, in the metric that contains values for load1, some of the ML nodes are missing. Only \%52.7 of the ML nodes are in the metric. The rest of the nodes and their values are lacking. There may be more missing nodes in various other metrics that we haven't investigated.
\\ \\ 
\textbf{The calculations while processing the data is wrong.}
    Even if we have checked the codes that are used to process the data, miscalculations might exist in the instrument and other script that we have made.
\\ \\
\textbf{ML workloads might differ from each other.}
    With the experience of the engineers at SURF, we label nodes in the LISA cluster having GPUs as machine learning nodes, because 90+\% of the workload belongs to the machine learning domain.
    The workload is run by various users, mainly scientists, and the ML workloads might be different from each other.

\section{Related Work} \label{sec:related}
There are several related works in the field for understanding how datacenters and other computer systems work. Although, there isn't any research on comparing datacenter operations during and before the coronavirus period. A similar study to our research is conducted by Uta et al. as the lack of open-source operational traces of low-level metrics is mitigated. The absence of such logs force large-scale systems experts and infrastructure developers to design, and implement their systems using unverified assumptions. The operational traces mitigates this issue and provides important information to system administrators and researchers on OS-level operations of datacenters. Furthermore, the traces enable performance evaluation, together with information on the meaning of the datacenter operations.


There are several related work conducted by experts in the field of workload data. Feitelson~\cite{feitelson2014experience} points out to the importance of workload data for evaluating performance in computer systems. The PWA contains 22 different logs, collected since 1993, and they contain accounting data about the jobs that executed on parallel supercomputers, clusters, and grids. The study also mentions about the quality issues of logs regarding missing data, inconsistent data, erroneous data an so on. 

\begin{figure}[H]
    \centering
    \includegraphics[width=\linewidth]{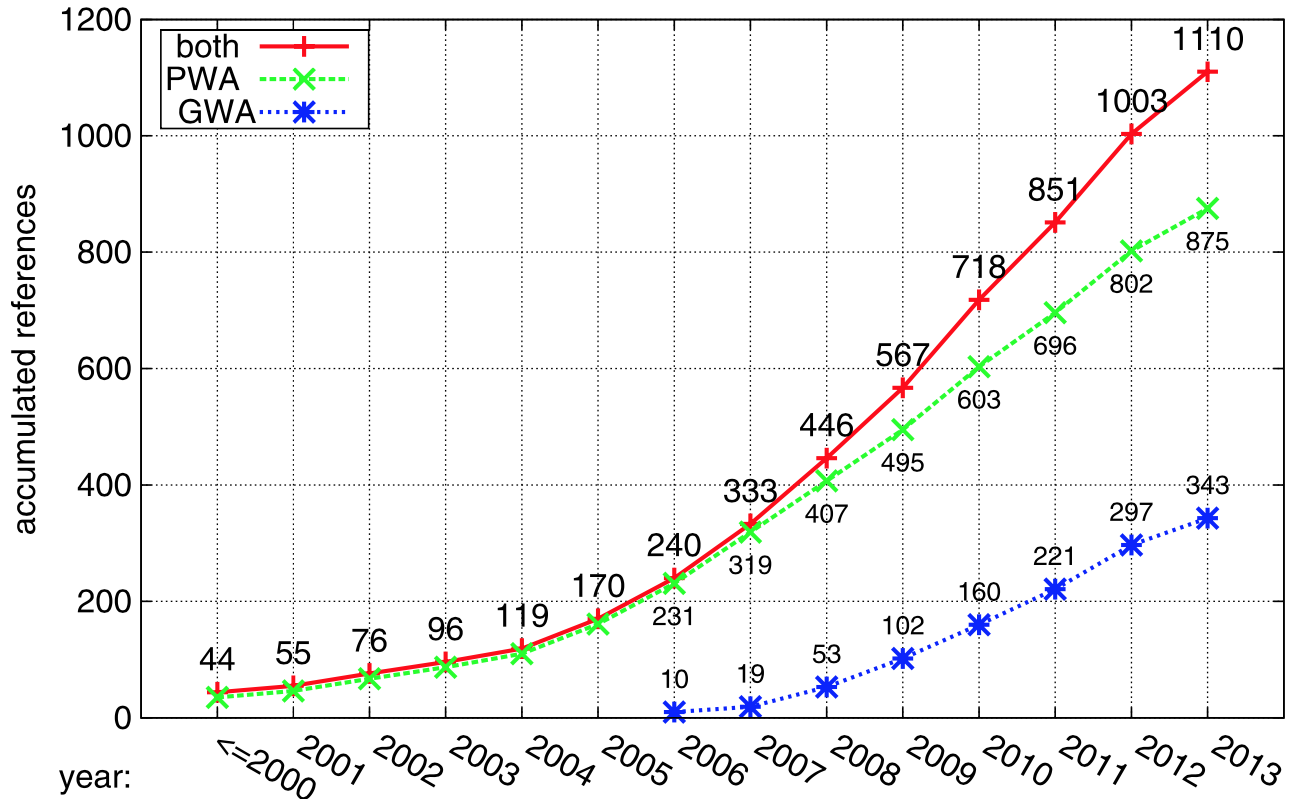}
    \caption{Accumulated yearly number of hits received when searching for the Parallel Workloads Archive (PWA) and the Grid Workloads Archive (GWA) in Google Scholar as of 28 October 2013~\cite{feitelson2014experience}}
    \label{fig:PWA_citations}
\end{figure}

According to Figure~\ref{fig:PWA_citations}, the PWA partially aids the need for workload data in the sense that the PWA has been cited more than 800 times since it was first open-sourced. In addition, both the GWA~\cite{iosup2008grid} and PWA have been cited more than 1100 times, which further proves the value of such data. However, due to certain regulations, the amount of workload traces are scarce and to address this issue, Iosup has published the GWA. The initial user group consists of researchers, industry, and education. In addition, for non-expert users, GWA has a mechanism for automated trace ranking and selection system so those users can also benefit and use the data for various purposes.

\begin{figure}[H]
    \centering
    \includegraphics[width=\linewidth]{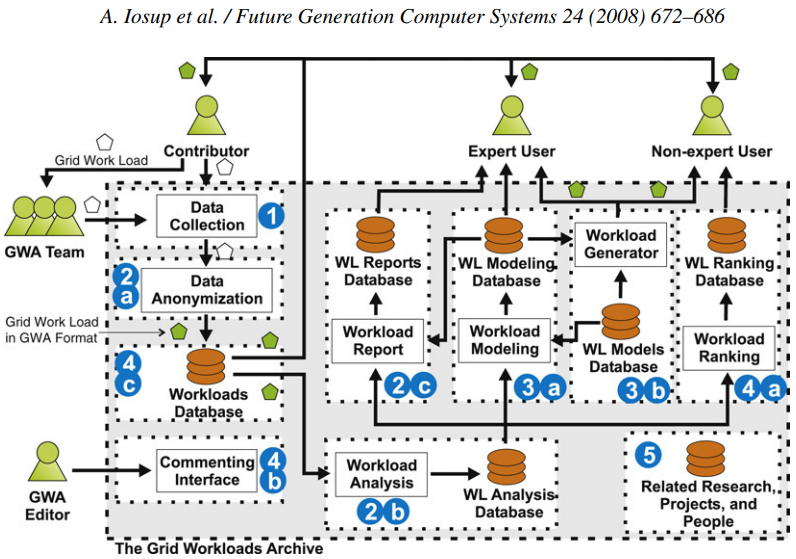}
    \caption{An overview of the Grid Workloads Archive design. The arrows represent the direction of data flows.~\cite{iosup2008grid}}
    \label{fig:gwa_design}
\end{figure}

In Figure~\ref{fig:gwa_design}, Iosup et al. shows the design requirements of the GWA. Box number one represents tools for collecting grid workloads, boxes with number 2 represent the tools for processing grid workload, Boxes with number 3 represent the tools for using grid workloads, and boxes with number 4 represent the tools for sharing grid workloads. After sharing grid workloads, they can be used for performance evaluations in computer systems, which is represented in box number 5. We can see that the GWA is a massive contribution to the computer systems community. It enables for unbiased research using real-world data. The results obtained from the data can be used to enhance grid computing and computer systems in general. 

Amvrosiadis et al. states that the main goal is to point out that most of the prior work, more than 450 publications, on trace analysis and characterization were done using the Google cluster traces~\cite{wilkes2011more}. To generalize the analysis and provide new traces, Amvrosiadis makes two contributions. First, he releases four traces: two from the private cluster Two Sigma, and two from HPC clusters located at the Los Alamos National Laboratory. Second, he examines the generality of workload characteristics obtained from the Google cluster trace. The findings in his research is that conducting new research using the Google trace alone is insufficient to guarantee generality. Therefore, this research is imperative in the sense that workload generality is important to understand how different clusters work.

\begin{figure}[H]
    \centering
    \includegraphics[width=\linewidth]{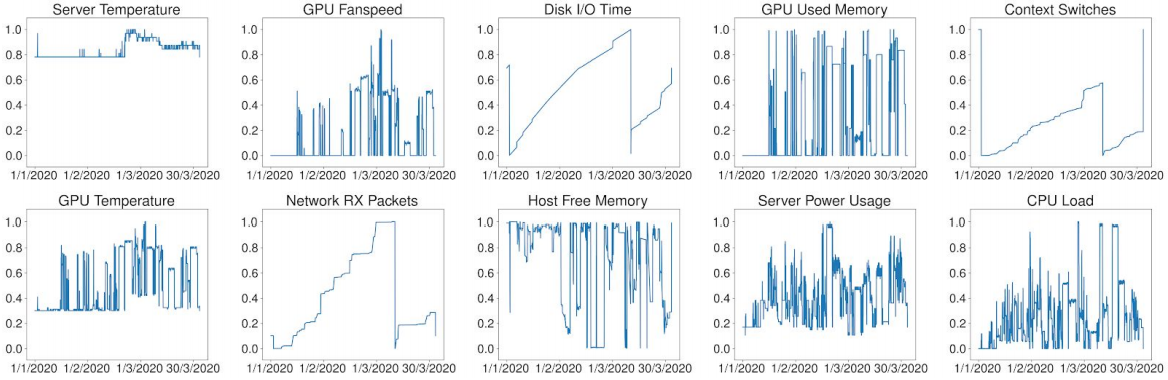}
    \caption{A dashboard to visualize 10 metrics for a single GPU-enabled server in the LISA cluster. Each metric is normalized by the maximum value encountered during the 3 months recorded for this server~\cite{alex_uta_2020}.}
    \label{fig:lisa_dashboard}
\end{figure}

Moreover, Uta et al. mentions that many analysis can be made using the open-sourced low-level traces. For instance, in Figure~\ref{fig:lisa_dashboard}, we can see a typical sysadmin dashboard from utilization-level metrics. Sysadmins can use these metrics to pinpoint an error or an abnormal situation in the system. Moreover, Uta advocates that using the operation logs, we can make systems more efficient, understand how certain workloads interact with the hardware, originate the errors, and gain deeper insights on the working of large-scale systems (datacenters). We can see that low-level operational logs important for various stakeholders. The main contribution Uta makes is open-sourcing fine-grained low-level operational logs spanning 3 months. Another contribution is a schema used for the collection of the datacenter metrics. This section demonstrates the importance of operational traces. Using the right traces, we can understand how a cluster or the whole datacenter works. The operations of a cluster can tell us a lot about what can we improve.

\section{Conclusion and Future Work} \label{sec:conclusion}
The coronavirus crisis has changed the way we work. Before the covid period, people mostly went to their offices work. During the global pandemic, people work from home and use online resources like zoom and google meet more. With the increased demand for datacenters, the operations of clusters might have changed during the covid period. In this research, we analyzed traces from a scientific cluster~(LISA) in the SURF datacenter. The process of extracting and analyzing these traces is explained in Section~\ref{sec:methodology}. We designed and developed a scientific instrument that can statistically compare two different time periods. The instrument produced graphs for the covid and non-covid period. We have seen that the power consumption, CPU usage, and temperature has increased during the covid period. The ram utilization was higher during the non-covid period. Moreover, the daily and hourly graphs showed us that the load1 values and power consumption has increased for the Generic nodes. The graphs that contain rack level utilization depicted that almost all the racks consumed more power, had higher temperature, and utilized their CPU more during the covid period. The instrument also produced node level curves with specified custom periods, which gives system administrators and data engineers the freedom to investigate any node for any given period. 

The traces we used in this research are the most fine-grained traces so far. Future work can further analyze these traces. Hopefully, after the covid period ends, traces can be extracted and analyzed comparing the pre-covid, covid and post-covid periods. 
\appendix

\section{Problem Statement} \label{app:problemstatement}

Due to coronavirus, many countries have made regulations on traveling and lockdowns have been issued. After the coronavirus lockdowns, the demand for services(e.g., video conferencing apps, social media, educational platforms, online gaming, streaming media) offered by datacenters has arisen and this has affected the operations of datacenters in an unprecedented way. Consequently, during the coronavirus period, the datacenter operations may have changed in a way that pressured datacenters due to intense workloads and understanding the change is in interest.

In order to understand the change and compare the difference of workloads, we need empirical, real-world data which is unavailable at the moment. Thus, we first need a tool to collect such data. Afterwards, we need to combine the collected data for meaningful analysis. Hence, a well designed instrument is required to accomplish all of the tasks (collect-combine-analyze workload data). Luckily, Uta et al.~\cite{alex_uta_2020} has collected and open-sourced workload data from SURFsara that has the finest-granularity out of all the other traces published so far. However, the instrument for combining and analyzing workload data is lacking. Hence, we need an instrument that can combine and analyze the collected workload data. The tool and instrument to analyze the datacenter workload are required to find out how datacenter operations have changed regarding workloads. To find the meaning of datacenter workloads in certain time scales(periods), an instrument will be designed that statistically compares the workload of a datacenter across different time periods. Using the designed instrument, we will check if there are any significant differences between the datacenter operations pre and during the coronavirus period.

\section{Self-Reflection} \label{app:selfreflection}
This research has given me the opportunity learn how datacenters work in terms of low-level operations. I learned how to process and analyze data. I learned how to design and develop a scientific instrument. I worked a lot on this research project, mostly part-time. Most of the time was spent on developing the instrument. Obtaining meaningful results and comparing graphs, tables were also challenging. Together with interpreting the selected graphs, selecting those graphs for different metrics was also time consuming. A significant amount of time was also spent on reading other related papers on datacenter workload characterization. In the last phase of this research, I got the coronavirus and I was sick for 3-4 days. Perhaps, the covid period didn't like being analyzed.

\bibliographystyle{abbrv}
\bibliography{main}

\end{document}